\documentclass[prl,twocolumn,showpacs,showkeys,preprintnumbers,floats,floatfix]{revtex4}

\usepackage{graphicx}

\begin{document}

\title{Broadband electrically detected magnetic resonance of phosphorus donors\\ in a silicon field-effect transistor}

\author{L.H. Willems van Beveren}
\email[Electronic mail: ]{l.h.willemsvanbeveren@unsw.edu.au}
\affiliation{Australian Research Council Centre of Excellence for Quantum Computer Technology, The University of New South Wales, Sydney 2052, Australia}
\author{H. Huebl}
\affiliation{Australian Research Council Centre of Excellence for Quantum Computer Technology, The University of New South Wales, Sydney 2052, Australia}
\author{D.R. McCamey}
\altaffiliation[ present address:~]{Department of Physics, University of Utah, 115 S 1400 E, Suite 201, Salt Lake City, Utah 84112-0830, USA}
\affiliation{Australian Research Council Centre of Excellence for Quantum Computer Technology, The University of New South Wales, Sydney 2052, Australia}
\author{T. Duty}
\altaffiliation[ present address:~]{School of Physical Sciences, The University of Queensland, St. Lucia 4072, Australia}
\affiliation{Australian Research Council Centre of Excellence for Quantum Computer Technology, The University of New South Wales, Sydney 2052, Australia}
\author{A.J. Ferguson}
\altaffiliation[ present address:~]{Cavendish Laboratory, JJ Thomson Avenue, Cambridge CB3 0HE, UK\\}
\affiliation{Australian Research Council Centre of Excellence for Quantum Computer Technology, The University of New South Wales, Sydney 2052, Australia}
\author{M.S. Brandt}
\affiliation{Walter Schottky Institute, Technische Universit\"{a}t M\"{u}nchen, Am Coulombwall 3, D-85748 Garching, Germany}
\author{R.G. Clark}
\affiliation{Australian Research Council Centre of Excellence for Quantum Computer Technology, The University of New South Wales, Sydney 2052, Australia}

\date{\today}

\begin{abstract}
We report electrically detected magnetic resonance of phosphorus donors in a silicon field-effect-transistor. An on-chip transmission line is used to generate the oscillating magnetic field allowing broadband operation. At milli-kelvin temperatures, continuous wave spectra were obtained up to 40 GHz, using both magnetic field and microwave frequency modulation. The spectra reveal the hyperfine-split electron spin resonances characteristic for Si:P and a central feature which displays the fingerprint of spin-spin scattering in the two-dimensional electron gas.
\end{abstract}

\pacs{71.55.-i, 73.20.-r, 76.30.-v, 84.40.Az, 85.40.Ry}

\keywords{EDMR, MOSFET, Si:P, ESR, stripline, $g$-factor, hyperfine}

\maketitle

In a semiconductor, conduction-band electrons and neutral, paramagnetic impurities have a spin-dependent scattering cross-section as demonstrated in silicon metal-oxide-semiconductor field-effect transistors (MOSFETs) by Ghosh and Silsbee~\cite{ghosh1992}. A resonant change of the source-drain current was observed at liquid-helium temperatures originating from the excitation of the electronic spin system. The electrically detected magnetic resonance (EDMR) spectra showed electron spin resonance (ESR) features with a hyperfine (HF) splitting of 4.2 mT and a central feature at $g$=2.000$\pm$0.001 which was at least partially attributed to conduction electrons. More recently, EDMR of neutral antimony donors was reported in an accumulation-mode field-effect transistor~\cite{lo2007}. Here, for a cavity frequency of 9.6 GHz and a temperature of 5~K, the six-fold HF-splitting of the $^{121}$Sb donors was observed, together with a gate voltage-dependent feature at $g$=1.9998, associated with the spin resonance of the two-dimensional electron gas (2DEG)~\cite{graeff1999,jantsch1998}.

In this letter, we report EDMR spectroscopy in P-doped silicon MOSFETs at milli-kelvin temperatures. A short-circuited non-resonant coplanar stripline (CPS)~\cite{koppens2006} was used to generate the on-chip ESR field and act simultaneously as the gate electrode. In contrast to previously reported resonator-based EDMR experiments, this allows broadband operation. Apart from the two Si:P HF lines, we observe a third line centrally positioned between the two P resonances, which we also attribute to the spin-dependent response of the 2DEG.

\begin{figure}[t]
\includegraphics[width=5.0cm]{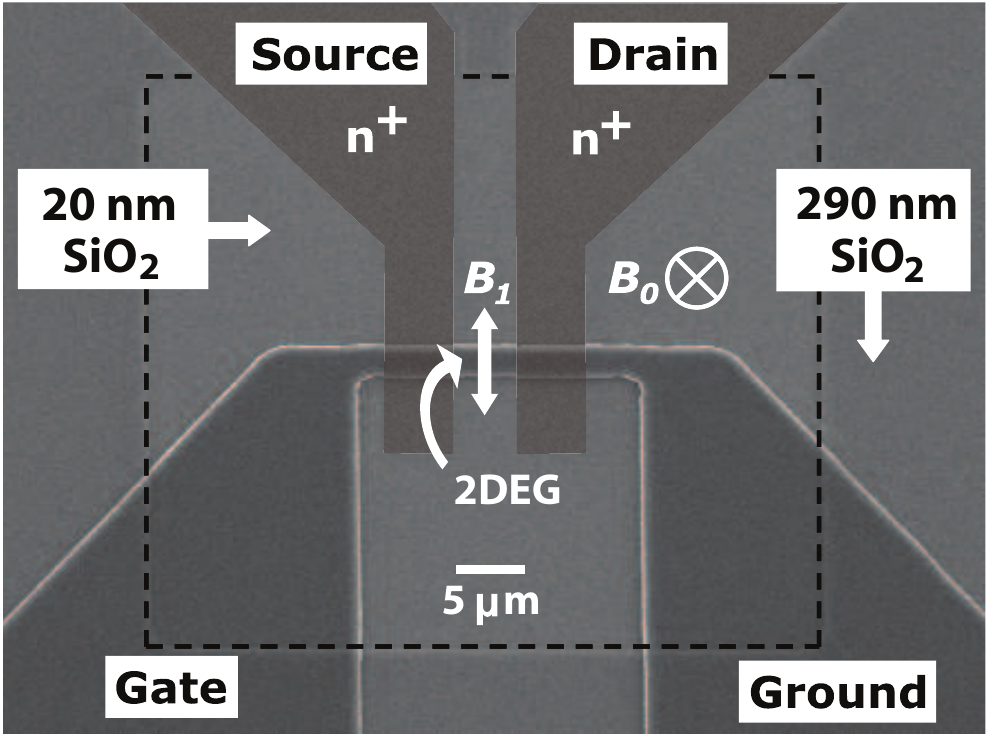}
\caption{Scanning electron micrograph of the MOSFET. Indicated are the n$^{+}$ source-drain contacts, the thin oxide window and the aluminum gate. The shorted CPS provides an anti-node in the current and therefore a maximum magnetic field.}
\label{fig1}
\end{figure}

The MOSFETs were fabricated photolithographically on bulk-doped silicon (100) with natural isotope composition. The room temperature resistivity of the substrates was 0.1 $\Omega$ cm, corresponding to a concentration of [P]=8$\times10^{16}$/cm$^{3}$. Firstly, a field oxide was grown to mask the P diffusion that defines the source and drain contacts (cf. Fig.~\ref{fig1}). Secondly, after local etch back of the field oxide, a high quality gate oxide was grown by ultra-dry oxidation. Finally, a 250 nm thick aluminum film was deposited to form the MOSFET gate and CPS. The channel area between the ohmic contacts is 5.0$\times$2.5 $\mu$m$^{2}$ and contains $\sim$10$^{4}$ P atoms assuming the accumulation layer extends to a depth of 10 nm below the Si/SiO$_{2}$ interface~\cite{ando1982}.

The MOSFET was turned on by lowering both the source and drain potentials with respect to the gate. The effective gate voltage $V_{G}$ is defined as the potential difference between the gate and the drain contact. While the induced accumulation layer carries an electrical current from source to drain, a microwave (MW) signal is applied to the CPS. The resulting oscillating magnetic field $B_{1}$ is oriented perpendicular to the static magnetic field $B_{0}$ which is generated by a superconducting magnet located in the dewar of a dilution refrigerator. To obtain the EDMR spectra, $B_{0}$ was changed in steps of 0.05 mT. To detect the change in the current, a current-to-voltage converter and bandpass filters were employed, in combination with a lock-in amplifier. Either magnetic field modulation (with a peak-to-peak modulation amplitude of $\sim$0.05 mT) or microwave frequency modulation (peak-to-peak modulation depth of 6 MHz) was used for phase-sensitive detection. A smaller superconducting magnet placed directly beneath the sample was used for magnetic field modulation. The MW power applied at the source was typically $P$=10-100 mW for frequencies in the range of $f$=20-40 GHz. The MOSFET was characterized at liquid-helium temperatures, where the substrate conductivity is negligible. The current flow through the MOSFET was found to be strongly enhanced by the MW signal. The MOSFET was operated in the linear regime where the 2DEG density is uniform throughout the channel and linearly proportional to the gate voltage.

\begin{figure}[t]
\includegraphics[width=5.0cm]{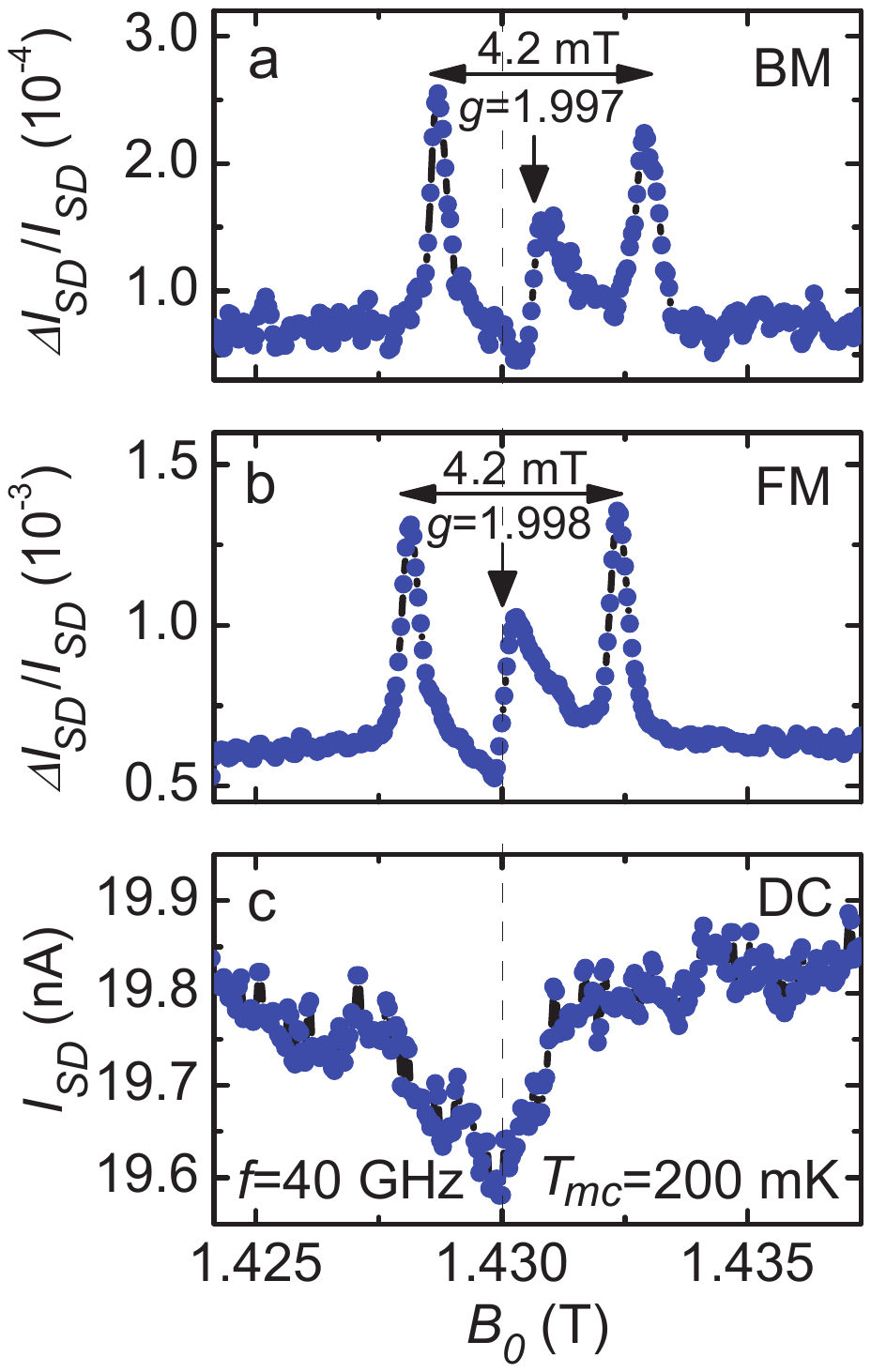}
\caption{(Color online) (a) EDMR spectrum (4 sweeps) obtained by magnetic field modulation (BM) at $f$=40 GHz using $P$=100 mW. The MOSFET was operated at $V_{G}$=150 mV and $V_{SD}$=100 mV, resulting in $I_{SD}$=160.14 nA. (b) EDMR spectrum (16 sweeps) obtained by frequency modulation (FM) using $P$=10 mW. For $V_{G}$=144 mV and $V_{SD}$=100 mV, the resulting average current was $I_{SD}$=19.76 nA. (c) Magnetic field dependence of $I_{SD}$, measured simultaneously with the FM spectrum in (b).}
\label{fig2}
\end{figure}

The EDMR spectra at $f$=40 GHz obtained by magnetic field modulation (BM) and microwave frequency modulation (FM) are shown in Fig.~\ref{fig2}(a) and Fig.~\ref{fig2}(b) respectively. Both spectra were obtained at a modulation frequency $f_{mod}$=13 Hz. Here, the resonant current change $\Delta I_{SD}$ detected by the lock-in amplifier is normalized to the average source-drain current $I_{SD}$, which is measured simultaneously. Under these conditions, the temperature of the mixing chamber was $T_{mc}$=200 mK. The EDMR spectra obtained by BM and FM have identical lineshapes, but the FM spectrum has a reduced noise floor which can be understood by the larger number of averages and the higher modulation depth for FM, corresponding to $\sim$0.2 mT in terms of magnetic field modulation amplitude. The EDMR signal amplitude in Fig.~\ref{fig2}(b) is larger compared to Fig.~\ref{fig2}(a) even though a smaller MW power was used. This can be explained by the $I_{SD}$ dependence of the EDMR signal amplitude, see Fig.~\ref{fig4}. The FM spectrum is shifted 0.55 mT with respect to the BM spectrum. This shift is attributed to long term drifts in the magnetic field.

Both spectra contain three resonances, where the resonances at $B_{0}$=1.4287 and $B_{0}$=1.4329 T (BM) correspond to the signature of isolated P donors in silicon with a characteristic hyperfine splitting of 4.2 mT. The center-of-gravity of the HF lines lies at $B_{0}$=1.4308 T, corresponding to a $g$-factor of $g_{P}$=1.9974 (1.9982 for FM). The central resonance at $B_{0}$=1.43065 T corresponds to a $g$-factor of $g_{c}$=1.9976 (1.99847 for FM) and is attributed to the spin-dependent electron-electron signal of the 2DEG as discussed later in this paper. The observed lineshape of the P donors, a non-derivative Lorentzian, is unexpected for magnetic field or frequency modulated ESR in slow adiabatic passage. Lineshapes similar to these have been seen in conventional ESR under fast adiabatic passage conditions, where the modulation period and sweep time is significantly shorter than the spin relaxation time $T_{1}$~\cite{cullis1975,weger1960}. This could also be the case here, because P electron spins have long relaxation times at liquid-helium temperatures~\cite{feher1959b}.

In contrast, the central resonance displays a bimodal or first-derivative lineshape. Although the spin relaxation of a 2DEG has only be studied at liquid-helium temperatures and for transfer-doped samples~\cite{graeff1999,tyryshkin2005}, the values of several microseconds observed there suggest a much faster relaxation compared to that in a dilute system of donors~\cite{tyryshkin2003}. The linewidth of the HF resonances is $\sim$0.5 mT, and the peak-to-peak width of the central resonance is 0.55 mT, both independent of MW frequency. None of the resonances are power broadened in the regime investigated here. The HF resonances are most likely broadened by the superhyperfine interaction with the non-zero $^{29}$Si nuclear spins in the crystal~\cite{feher1959a}.

Figure~\ref{fig2}(c) shows $I_{SD}$ (DC) while sweeping through the resonances. A decrease was observed which is most pronounced at the central resonance. This indicates that saturation of the electron spin transitions of the 2DEG enhances the resistivity of the MOSFET channel as seen by the conduction-band electrons.

\begin{figure}[t]
\includegraphics[width=5.0cm]{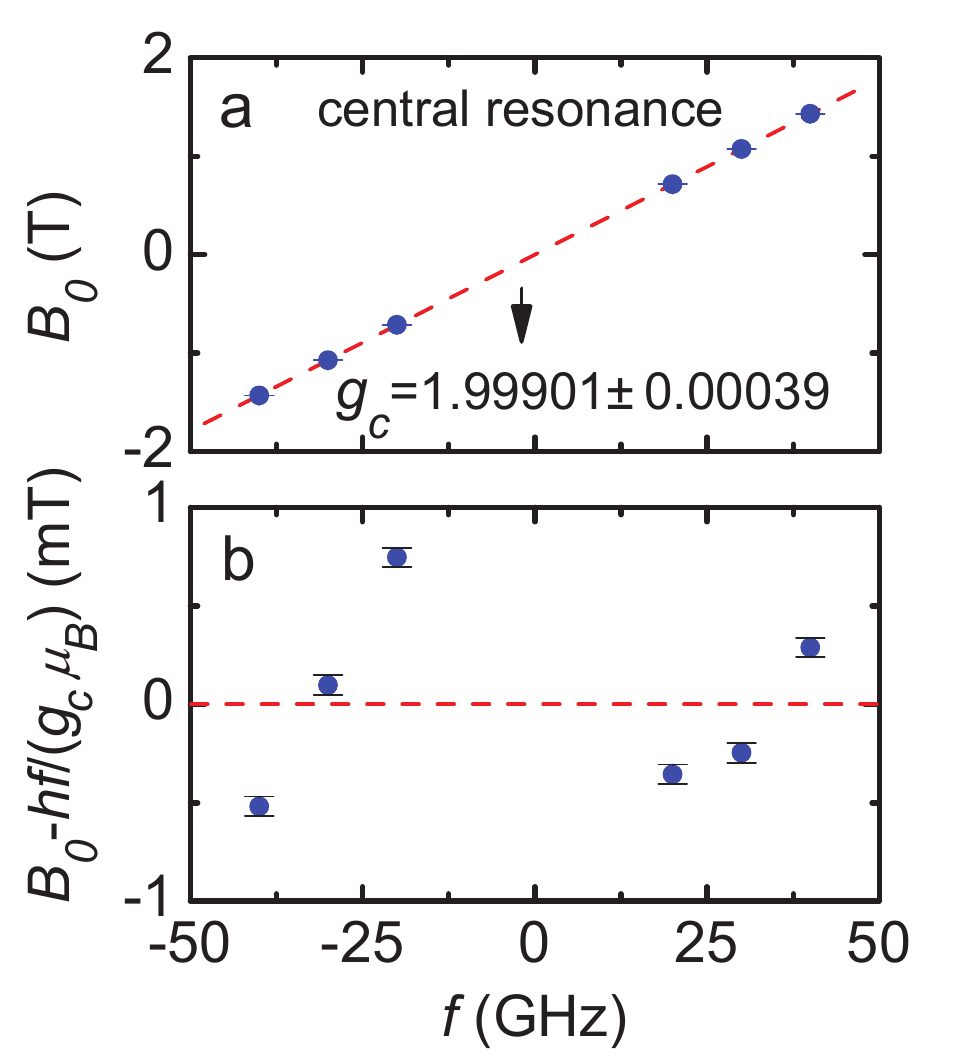}
\caption{(Color online) (a) Resonance fields of the central line as a function of MW frequency. (b) Residual data plot.}
\label{fig3}
\end{figure}

The CPS allows EDMR spectroscopy in the frequency range $f$=20-40 GHz, as shown in Fig.~\ref{fig3}(a). The corresponding spectra were taken with frequency modulation at $f_{mod}$=10.123 kHz. Plotting the resonance fields as a function of MW frequency allows an accurate determination of the $g$-factor, even in the presence of an offset magnetic field, by linearly fitting the data according to $hf$=$g\mu_{B}B_{0}$. Figure~\ref{fig3}(b) depicts the residual plot of the data points using the linear fit of Fig.~\ref{fig3}(a). For the central resonance we find $g_{c}$=1.99901$\pm$0.00039. For a donor impurity band, a $g$-factor of 1.99875 has been reported, while electrons in the conduction band of bulk silicon have a $g$-factor of 1.9995~\cite{young1997}, which increases to $g$=2.0000 in 2DEGs~\cite{graeff1999,jantsch1998}. Probably due to the limited mobility of the 2DEG, the resonance observed here is considerably broader than those reported for bulk~\cite{young1997} or transfer-doped 2DEGs~\cite{graeff1999,jantsch1998}, but it is in good agreement with the width reported for a similar structure~\cite{ghosh1992,lo2007}, in contrast to the linewidth of 0.06 mT reported more recently~\cite{shankar2008}. For the center-of-gravity of the HF lines we find $g_{P}$=1.99874$\pm$0.00036, in agreement with the value of $g_{P}$=1.99850 obtained by conventional ESR~\cite{young1997}.

Figure~\ref{fig4} shows $\Delta I_{SD}/I_{SD}$ for the central resonance (peak-to-peak) and the HF lines (base-to-peak) as a function of $I_{SD}$ by varying $V_{G}$. The corresponding EDMR spectra were obtained by magnetic field modulation. At high $I_{SD}$ the signal intensity for the central resonance shows a $I_{SD}^{-2}$ dependence. If we assume a $V_{G}$-independent mobility, $I_{SD}$ is directly proportional to the 2DEG density $n_{e}$. The $n_{e}^{-2}$ dependence of the central resonance is the fingerprint for electron-electron scattering in the 2DEG as the dominant spin-dependent process~\cite{ghosh1992,graeff1999}. For the HF resonances a $I_{SD}^{-1/2}$ dependence is observed. If scattering between the 2DEG and the P donors gives rise to these EDMR resonances, the signal intensity would be proportional to $P_{P}P_{n_{e}}$~\cite{lepine1972}, where $P_{P}$=$\tanh(g_{P}\mu_{B} B_{0}/2k_{B}T)$ is the thermal polarization of the ensemble of P donors and $P_{n_{e}}\propto n_{e}^{-1}$ the polarization of the 2DEG, resulting in $\Delta I_{SD}/I_{SD}\propto n_{e}^{-1}\propto I_{SD}^{-1}$.

\begin{figure}[t]
\includegraphics[width=5.0cm]{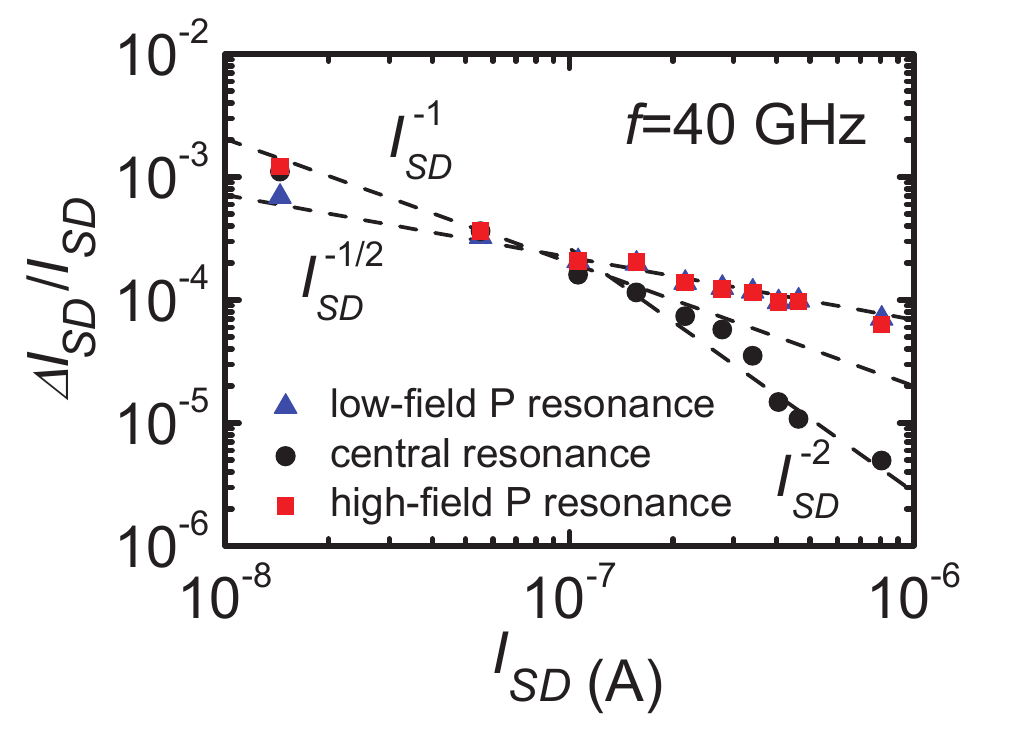}
\caption{(Color online) Normalized EDMR signal intensity for the central resonance and HF lines as a function of $I_{SD}$.}
\label{fig4}
\end{figure}

This work demonstrates EDMR of P donors in a silicon MOSFET at milli-kelvin temperatures, where the spin polarization of the P donors is expected to be very high. The broadband setup allowed phase-sensitive detection using both magnetic field and frequency modulation, resulting in quantitatively equivalent EDMR spectra. The Lorentzian lineshape of the HF resonances could indicate a very long $T_{1}$ time for Si:P. The broadband capability of the spectrometer also allowed a precise determination of the $g$-factor of all resonances. The $I_{SD}^{-2}$ dependence of the central resonance indicates spin-spin scattering between electrons in the 2DEG. This local ESR technique could be used to implement single-spin manipulation and control applicable, for example, in the Kane proposal for silicon-based quantum computing~\cite{kane1998}.

The authors would like to thank R.P. Starrett, D. Barber and E. Gauja for technical assistance. This work is funded by the Australian Research Council, the Australian Government, the U.S. National Security Agency and the U.S. Army Research Office (under Contract No. W911NF-04-1-0290). H.H. and M.S.B. would like to thank the Deutsche Forschungsgemeinschaft for financial support (SFB631).

\end{document}